\documentclass{ws-p8-50x6-00}

\begin{document}

\title{On Tachyon kinks from the DBI action }

\author{Ph.Brax}

\address{\  Service de Physique Th\'eorique,
CEA Saclay, 91191
Gif-sur-Yvette, France}

\author{J.Mourad and D.A.Steer}

\address{
Laboratoire de Physique Th\'eorique\footnote{Unit\'e Mixte de
Recherche du CNRS (UMR 8627).}, B\^at. 210, Universit\'e
Paris XI, \\ 91405 Orsay Cedex, France\\
and
\\
F\'ed\'eration de recherche APC, Universit\'e Paris VII,\\
2 place Jussieu - 75251 Paris Cedex 05, France.
}


\maketitle

\abstracts{We consider solitonic solutions of the DBI tachyon effective
action for a non-BPS brane in the presence of an electric field.
We find that for a constant electric field $\tilde E\le 1$, regular
solitons compactified on a circle admit a singular and
decompactified limit corresponding to Sen's proposal provided the
tachyon potential satisfies some restrictions. On the other hand
for the critical electric field $\tilde E=1$, regular and finite energy
solitons are constructed without any restriction on the potential.
}

\section{Introduction}

In addition to   the stable D-branes, type II superstrings admit
non-supersymmetric D-branes\cite{sen}. The instability of these
non-BPS branes is signalled by the presence of a tachyon on their
worldvolume, and their decay is described by the dynamics of the
tachyon.
 The BPS branes can be viewed as tachyon kinks on the
 non-BPS branes with one dimension higher\cite{sen2}.
 The dynamics of the decaying tachyon can be
 captured\cite{Sen:1999md,Garousi:2000tr,Bergshoeff:2000dq,Kluson:2000iy,Sen:2002an,Sen1,Kutasov:2003er,Garousi:2003pv}
using the Dirac-Born-Infeld (DBI):
 \be
 S = - \int d^px \, dt \, V(T)
 \sqrt{\left( 1 +  \partial_\mu T \partial_\nu
T \eta^{\mu \nu} \right)}. \label{dbi} \ee Here $V(T)$ is the
tachyon potential which is even and vanishes at infinity where it
reaches its minimum.  Near the global maximum at $T=0$,
$V(T)={\cal T}_p(1-\beta^2 T^2/2)+\dots$ where ${\cal T}_p$ is the
tension of the non-BPS $p$-brane and the potential encodes the
mass of the tachyon near the perturbative vacuum $T=0$.  Finally
$\eta^{\mu \nu} = (-1,+1,\ldots,+1)$ is the $p+1$ Minkowski
metric.

It happens that finite energy solitons of the DBI action are
singular\cite{Sen:2003tm,Minahan:2000tg,Lambert:2002hk}. One would
like to regularise the behaviour of such solitons. This can be
achieved by compactifying one spatial dimension on a circle. The
singular solitons can be then obtained in the decompactifed limit
provided some restrictions on the tachyon potential are imposed.
On the other hand, there seems to be an intrinsic ambiguity in the
regularization process of the DBI action leading to singular
solutions. In particular, adding an extra kinetic term for the
tachyon and then taking the limit where this term
vanishes\footnote{Equivalently one can work with the action (2)
below with $q \neq 1/2$ and then take the limit $q \rightarrow
1/2^+$ at the end of the calculation.}  leads to singular kinks
with no restriction on the shape of the potential\cite{dani}. Such
an ambiguity can be lifted for kinks in an electric field
background. In particular for a critical value $\tilde{E}=1$ of
the electric field, one always finds regular and finite energy
solutions of the DBI action.

The plan of this paper is as follows. In Section 2 we revisit
Derrick's theorem which gives necessary conditions for the
existence of finite energy solitons for actions of the form
(\ref{dbi}). In Section 3, we compactify one spatial coordinate on
a circle of radius $R$ in order to get regular kinks. In fact, we
obtain regular solitons describing $n$ pairs of kinks and
anti-kinks. Then we determine the conditions on the potential for
which the decompactifying limit  exists. The stability of the
kink-antikink configurations is also described. Finally we
consider the case of charged kinks\cite{Sen:2003bc} and focus on the critical case
where regular and finite energy solitons on the infinite line can
be found.

\section{Uncharged solitons and Derrick's theorem}

We are interested in kink-like solutions of (\ref{dbi}). These
kink solutions should represent the stable BPS $p-1$ branes into
which the non-BPS $p$-brane decays.  Before doing this explicitly,
it is worthwhile recalling Derrick's theorem\cite{Rajaraman:is}:
in the case of the usual Klein-Gordon action for scalar fields, it
tells us that finite energy static solitonic solutions on an
infinite space are only possible in (1+1)-dimensions.  Here will
draw similar conclusions starting from (\ref{dbi}).

Let us  consider a slightly more general action: \be S = - \int
d^px \, dt \, V(T) \left( 1 +
\partial_\mu T \partial_\nu T \eta^{\mu \nu} \right)^q,
\label{actiongen} \ee where the real scalar field $T$ is
dimensionless (we set $\alpha'=1$ throughout). For $q=1/2$ this is
just (\ref{dbi}).
 Let $T_1(x)$ be a static solution of the equations of motion,
and hence an extremum, $\delta E=0$, of the finite static energy
functional
\be E[T] = \int d^p x V(T)  \left( 1 + \partial_i T \,
\partial^i T \right)^q.  \label{Etach} \ee
Now consider a family $T_\lambda (x)= T_1 (\lambda x)$, using
$\frac{d}{d\lambda} E[T_\lambda]\vert_{\lambda =1}=0$ we find
 \be
 \int d^p x V(T) \left( 1 +
\partial_i T \, \partial^i T \right)^{q-1} \left[  p+ (p-2q)
\partial_i T \,\partial^i T \right] =0. \label{derrickT} \ee
When $q=1/2$, the case we focus on here, the square bracket in
(\ref{derrickT}) is $p+(p-1) \partial_i T \partial^i T$ which can
never vanish for $p \geq 1$. Thus no finite energy static
solutions seem to be permitted.  However, there is formally a way
around this: when $p=1$ Eq.~(\ref{derrickT}) becomes \be 0=\int
d^p x \frac{V(T)}{ \sqrt{1 +  T'^2 }} \label{derrickT2} \ee which
can vanish if
$$T' \rightarrow \pm \infty
\qquad {\rm or} \qquad T \rightarrow \pm \infty
$$
with $E[T]$ remaining finite.

 Let us suppose that the
equations of motion admit a solution with $T' \rightarrow \pm
\infty$ describing a {\it single} kink on the infinite line (below
we will find the conditions on $V(T)$ such that this is the case),
\be
T(x) = \lim_{C \rightarrow 0} \frac{x}{C} = \left\{
\begin{array}{cl}
\infty & x > 0 \\
0 & x=0. \\
-\infty & x < 0
\end{array}
\right. \label{SenT}
\ee
This is a typical case of the solutions discussed by
Sen\cite{Sen:2003tm}: it describes an infinitely thin topological
kink interpolating between the two vacuua. Notice that $V=0$
everywhere apart from when $T=0=x$. Substitution of (\ref{SenT})
into the energy functional (\ref{Etach}) (taking carefully the $C
\rightarrow 0$ limit) gives the the energy of this singular
solution to be
\be
E_{\rm Sen} = \int_{-\infty}^{\infty} V(x) dx.
\label{energythin}
\ee
Sen has argued that such singular kinks are stable, as the BPS
brane should be, and furthermore that their effective action is
exactly the required DBI action\cite{Sen:2003tm}.  Hence such
solutions are of great interest, and would suggest that BPS branes
are infinitely thin.  Note that the parameters of $V(T)$ should be
tuned such that $E_{\rm Sen} = {\cal T}_{p-1}$.

In this note, we will construct such singular solitons  as limits
of  regularised kink solutions. Sen's singular limit can then be
approached in the decompactification limit. We are looking for
static kink-like solutions in which $T$ has a non-trivial
dependence on only  one spatial coordinate; $T=T(x)$. Let us
assume that the kink is centered at the origin, $T(0)=0$. The
equations of motion coming from (\ref{actiongen}) with $q=1/2$
have a first integral, \be {V(T) \over \sqrt{1+(T')^2}}=V_0,
\label{v0} \ee where $V_0 \geq 0$ is a constant. The energy of
this solution is given by \be E=\int dx V(T)\sqrt{1+(T')^2} =
\frac{1}{V_0} \int dx V^2(T). \label{E} \ee Since $T'^2$ is
positive, solutions of equation (\ref{v0}) exist in the region $
V(T) \geq V_0$. Furthermore, the solutions $T(x)$ are periodic
with a $V_0$-dependent amplitude which, from (\ref{v0}), diverges
as $V_0 \rightarrow 0$. Note that within one period there must be
both a kink and an anti-kink (corresponding to the two points for
which $T(x)=0$).  Also, the energy density of the kinks becomes
more and more localised as $V_0 \rightarrow 0$. Below we will
study in detail the dependence of both the period and $E$ on
$V_0$.

In order for $E$ to be finite, one must have $|T| \rightarrow
\infty$ as ${|x| \rightarrow \infty}$. This immediately implies
from (\ref{v0}) that $V_0 = 0$
--- a topological kink.\footnote{Notice that this condition is
identical to (\ref{derrickT2}).}
 Sen's solution\cite{Sen:2003tm} (\ref{SenT}) is such a
singular solution in which $T$ vanishes for only one value
of $x$: in other words Sen's solution is periodic with a divergent
wavelength.  Furthermore in that case $E = 2E_{\rm Sen}$.

We would like to approach the singular ($V_0 \rightarrow 0$)
kink(s) as the limit of regular solutions with $V_0 \neq 0$.  In
order to get regular and finite energy solutions we shall suppose
that $x$ is a compact direction of length $2\pi R$, so that $V_0$
can now be non-zero. Our aim is to see whether the limit
$R\rightarrow\infty$ with $E$ being finite exists. Note that if
this limit exists we expect to get twice (\ref{energythin}) as the
energy of the resulting solution. In fact,
 BPS branes have a RR charge and on a circle the
sum of the charges must vanish so all we can get are pairs of
branes and anti-branes. Sen's solution corresponds to the brane
infinitely distant from the anti-brane.

The period of the solutions of equation (\ref{v0}) is $4\zeta$
where \be \zeta(V_0) =\int_{0}^{T_0}{dT \over {\sqrt{\left(V\over
V_0\right)^2-1}}}, \label{zeta} \ee and $T_0$ is defined by
$V(T_0)=V_0$. The radius is thus given by
\be 2 \pi R=4n\zeta, \label{ra}\ee
 where $n=1,2,\dots$. The separation between the brane
and the anti-brane is $2\zeta$. The $V_0$-dependent energy of this
solution can be written as \be E(V_0) = 4n \int_{0}^{T_0}dT{ V
\over{\sqrt{1-\left( V_0\over V\right)^2}}} \label{ener} \ee or
equivalently
 $
4 n{\cal{E}}(V_0)$.

The behaviour of $\zeta(V_0)$ as $V_0 \rightarrow 0$ depends
critically on the form $V(T)$ for large $T$. Since the potential
is assumed positive, let us write it in the form
$V=e^{-\sigma(T)}$,
then the behavior of $\zeta(V_0)$ depends on $\sigma'(T)$
at large $T$ \cite{BMS}: $\zeta(V_0) \propto  [\sigma'(T_0)]^{-1}$.
Let us examine the limit where the radius goes to infinity. There
are three possibilities
\begin{enumerate}
\item
$\zeta \rightarrow \infty$ as ${V_0 \rightarrow 0}$.
This happens when $\sigma'\rightarrow 0$. The kink
anti-kink separation tends to infinity, and as $R \rightarrow
\infty$, equation (\ref{ra}) can be satisfied with $n=1$ (a single
kink and anti-kink).  The energy of this solution will, however,
depend on the behaviour of $\zeta V_0$.  Indeed,
the energy goes to
$2\int_{-\infty}^{\infty} V(T) dT$ if and only if $\zeta
V_0\rightarrow 0$ \cite{BMS}.
  This latter case is the one which reproduces a
single singular solution in the non-compact limit.

\item $\zeta \rightarrow$ const $\neq 0$ as ${V_0 \rightarrow 0}$.
Now the kink anti-kink distance tends to a constant and one can
never get an isolated kink.  As $R \rightarrow \infty$, equation
(\ref{ra}) can only be satisfied if $n \rightarrow \infty$.
In fact the realistic potential $V=v/\cosh(\beta T)$ falls in this
category.

\item $\zeta \rightarrow
0$ as $V_0\rightarrow 0$. The kink and anti-kink separation tends
to zero and it is clearly not possible to obtain a single kink and
anti-kink in the non-compact limit. This behaviour
occurs for many potentials considered in the literature; for
example $V(T) = v e^{-T^2}$.
\end{enumerate}


%

We now ask which  conditions $V(T)$ must satisfy so that as $V_
0$ and $R \rightarrow \infty$ the energy $E$ in
(\ref{ener}) with $n=1$ reduces to $2E_{\rm Sen}$ in
(\ref{energythin}). We find that  ${\cal{E}}(V_0)$ will be finite
if and only if, as $T\rightarrow \infty$, \be
 {|V'| \over V^2}\rightarrow \infty.
 \label{epsilon0}
 \ee
 In this case, ${\cal E} \rightarrow \int_0^{\infty} V(T)dT$.
Thus from (\ref{epsilon0}) we conclude that for exponential
potentials \be V(T \rightarrow \infty) \sim \exp(-T^{a}) \qquad
\Rightarrow \qquad {\cal{E}} \; {\rm finite} \; \forall \; a > 0
\label{conexpenergy} \ee whereas for power-law potentials \be V(T
\rightarrow \infty) \sim \frac{1}{T^{1/\alpha}} \qquad \Rightarrow
\qquad {\cal{E}} \; {\rm finite} \; \iff \; \alpha < 1.
\label{conpowerenergy}
\ee

 For any $R$ and non-zero $V_0$ the kink
anti-kink array is unstable\cite{BMS}. On the other hand the
singular kink is always stable.

\section{ Charged Solitons}

Now consider solitons in an electric field background 
\cite{Sen:2003bc} $A_0(x)$ in
the gauge where $A_1=0$ and all the other components of the gauge
field are set to zero. The DBI action depends on the tachyon and
the field strength $F_{\mu\nu}$:
\be
S=-\int d^{p+1} x V(T) \sqrt{-\gamma}
\ee
where $\gamma$ is the determinant of $\gamma_{\mu\nu}=
\eta_{\mu\nu} +\partial_\mu T
\partial_\nu T + F_{\mu\nu}$ and the only non-zero component of
$F_{\mu\nu}$ is $F_{01}=-\partial_1 A_0 (x)\equiv \tilde E$. The
dynamics are determined by the Lagrangian density
\be
{\cal L}=
V(T) \sqrt{ 1+ T'^2 -\tilde E^2}
\ee
which now has two second integrals $V_0$ and $e$ where
\be
V_0= \frac{V(T)}{\sqrt{1+T'^2 -\tilde E^2}}
\ee
and from the
Euler-Lagrange equations for $A_\mu$
\be
e=\frac{\tilde E V(T)} {\sqrt{1+T'^2 -\tilde E^2}}.
\ee
Since $V_0$ and $e$ are constant, the electric field
\be
\tilde E=\frac{e}{V_0}
\ee
is also constant. The tachyonic dynamics can be deduced from
\be \tilde V_0= \frac{V(T)}{\sqrt{ 1+ (\frac{dT}{dy})^2}} \ee
where $ dy =\sqrt {1-\tilde E^2} dx$ and
$\tilde V_0 = \sqrt{1-\tilde E^2} V_0$.
Notice that there is a maximal value for the electric field
$\tilde E_0=1$. As soon as $ \tilde E< \tilde E_0$,
one can use the previous analysis
on the shape of the potential $V(T)$ with the same restrictions as
in the neutral case.

More interestingly, the critical case $\tilde E=1$
leads to non-singular
solitons on the real line.
In that case the dynamics are governed
by \be T'^2 =\frac{V^2}{V_0^2}. \ee Let us define $T_{V_0}(x)=
T_1(\frac{x}{V_0})$ then $T_1$ satisfies the universal profile \be
\vert T_1'\vert = V(T).\ee The solution \be x= \int _0 ^T \frac{
du}{V(u)}\ee defines a soliton interpolating between the $
T=-\infty$ at $x=-\infty$ and $T=\infty$ at $x=\infty$, i.e.
between two vacua of the theory. The width of the soliton is of
order $O(V_0)$ implying that one gets an infinitely thin soliton
in the $V_0 \to 0$ limit. For any other value of $V_0$ the
soliton is of finite width.

Finally notice that for any value of $V_0$, the energy of the
solitons when $\tilde E=1$ is simply
\be
E=
\int_{-\infty}^{\infty} V(T) dT,
\ee
i.e.~Sen's value for the tension of the BPS brane. This value is
independent of the width $V_0$ of the soliton. Therefore the
zero-width limit does not modify the energy of the soliton.

We have thus found that introducing a constant electric field
allows one to define regular solitons of the non-BPS DBI action on
the infinite line. Moreover infinitely thin solitons can also be
obtained by taking the limit where the width of the regular
solitons goes to zero.

\section{Conclusion}

We have performed an analysis of the regular solitons on the
circle
of the non-BPS DBI action, both in the absence and in the presence
of a constant electric field in the soliton direction. For an
electric field below the critical value $\tilde{E}=1$ and with the
regularisation scheme presented here, one finds that singular
solitons on the infinite line can only be obtained for restricted
tachyon potentials. On the other hand when $\tilde E=1$, regular
solitons of finite energy  exist in noncompact space.

\end{document}